\documentclass[twocolumn,showpacs,floatfix,superscriptaddress,amsmath,amssymb,prb]{revtex4}
\usepackage{mathrsfs}
\usepackage{txfonts}
\usepackage{amssymb}
\usepackage{graphicx}
\usepackage{hyperref}
\usepackage{ulem}
 \usepackage{overpic}
 \usepackage{psfrag}
\usepackage{tabularx}
\usepackage{array}
\usepackage{placeins}

\begin{document}
\title{Long-range string orders and topological quantum phase transitions
        in the one-dimensional quantum compass model}

\author{Hai Tao Wang}
\affiliation{Centre for Modern Physics and Department of Physics,
Chongqing University, Chongqing 400044, The People's Republic of
China}
 \author{Sam Young Cho}
 \email{sycho@cqu.edu.cn}
\affiliation{Centre for Modern Physics and Department of Physics,
Chongqing University, Chongqing 400044, The People's Republic of
China}

\begin{abstract}
 In order to investigate the quantum phase transition in the one-dimensional quantum compass model,
 we numerically calculate non-local string correlations, entanglement entropy,
 and fidelity per lattice site by using
 the infinite matrix product state representation with the infinite time
 evolving block decimation method.
 In the whole range of the interaction parameters,
 we find that the four distinct string orders characterize the four different Haldane phases
 and the topological quantum phase transition occurs between the Haldane phases.
 The critical exponents of the string order parameters $\beta=1/8$ and the cental charges
 $c=1/2$ at the critical points show that the topological phase transitions between the phases belong to an Ising type of universality classes.
 In addition to the string order parameters,
 the singularities of the second derivative of the ground state energies per site, the continuous
 and singular behaviors of the von Neumann entropy, and the pinch points of the fidelity per lattice site manifest that
 the phase transitions between the phases are of the second-order,
 in contrast to the first-order transition suggested in pervious studies.
\end{abstract}

\maketitle

\section{Introduction}

 Transition metal oxides (TMOs)
 with orbital degeneracies
 have been intensively studied for quantum phase transitions (QPTs) because
 they have shown extremely rich phase diagrams due to
 competitions between orbital orderings and
 complex interplays between quantum fluctuations and spin interactions
~\cite{Brzezicki1,You1,Brzezicki2,Sun1,Eriksson,Mahdavifar1,Liu1,Sun2,
 Wang,Jafari1,Mahdavifar2,You2,Liu2,Wenzel,Orus,Jackeli}.
 In order to mimic such competitions between orbital ordering in different directions
 and directional natures of the orbital states with twofold degeneracy
 in the language of the pseudospin-$1/2$ operators,
 Kugel and Khomskii~\cite{Kugel} first introduced the quantum compass model (QCM) in 1973.
 In this model, the pseudospin-1/2 operators
 characterize the orbital degrees of freedom, and the anisotropic couplings between these pseudospins simulate the competition between orbital orderings in different directions.
 Furthermore, such an idea has been implemented to describe
 some Mott insulators with orbital degeneracy~\cite{Feiner,Dorier},
 polar molecules in optical lattices~\cite{Micheli} and ion trap systems~\cite{Milman},
 protected qubits for quantum computation in Josephson junction arrays~\cite{Doucot}, and so on.

 Based on the one-dimensional QCM,
 physical properties and QPTs in TMOs
 have been explored
 in the absence ~\cite{Brzezicki1,You1,Brzezicki2,Sun1,Eriksson,Mahdavifar1,Liu1}
 or in the presence ~\cite{Sun2,Wang,Jafari1,Mahdavifar2,You2,Liu2}
 of a transverse magnetic field.
 Especially for its criticality,
 in 2007, Brzezicki {\it et al.}~\cite{Brzezicki1}
 used the Jordan-Wigner transformation mapping it to an Ising model,
 obtained an exact solution of the QCM, and
 suggested that the system has the first-order transition occurring between two disordered phases.
 In 2008, You and Tian~\cite{You1} supported Brzezicki {\it et al.}'s result, i.e.,
 the first-order transition
 by adopting the reflection positivity technique
 in the standard pseudospin representation.
 In 2009, in order to show that the phase
 transition is intrinsic to the system not an artifact originating from a singular parameterization
 of the exchange interactions, Brzezicki and Ole\'{s}~\cite{Brzezicki2} revisited the model
 and reclaimed the first-order transition.
 In the same year, furthermore, Sun {\it et al.}~\cite{Sun1} reached to a conclusion supporting
 the first-order phase transition between two different disordered phases
 by using the fidelity susceptibility and the concurrence.
 However,
 Sun and Chen \cite{Sun2} considered a transverse magnetic field on the QCM and
 found at the zero field
 that the phase transition is of the second-order
 from the finite-size scaling of the spin-spin correction
 as well as the fidelity susceptibility,
 the block entanglement entropy, and the concurrence.
 Also,
 following in Ref.~\onlinecite{Brzezicki1} but introducing one more tunable parameter,
 Eriksson and Johannesson~\cite{Eriksson} noticed
 the second-order phase transition rather than the first-order phase transition by using
 the concurrence and the block entanglement at the multicritical point in the one-dimensional
 extended quantum compass model (EQCM).
 In 2012, Liu {\it et al.}~\cite{Liu1} numerically studied the  EQCM
 by utilizing the matrix product state (MPS) with
 the infinite time-evolving block decimation (iTEBD) algorithm
 and, at the multicritical point,
 observed both features of the first-order and
 the second-order phase transition.
 Thus, the phase transition in the one-dimensional QCM
 is still not characterized clearly.

%%%%%%%%%%%%%%%%%%%%%%%%%%%%%%%%%%%%%%%%%%%%%%%%%%%%%%%
\begin{figure}
\includegraphics [width=0.3\textwidth]{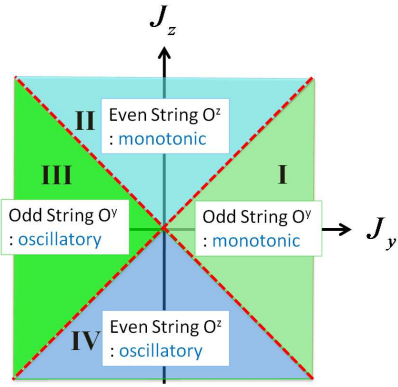}
\caption{ (Color online)
 Groundstate phase diagram for
 the one-dimensional QCM in $J_y$-$J_z$ plane.
 The four topologically ordered phases
 are characterized by the four distinct string orders (defined in the text).
 The critical lines are (i) $J_y=J_z > 0$ ($\theta=\pi/4$), (ii) $J_y=-J_z < 0$ ($\theta=3\pi/4$),
 (iii) $J_y=J_z < 0$ ($\theta=5\pi/4$), and (iv) $J_y=-J_z < 0$ ($\theta=7\pi/4$).
 At the critical points, the central charges are $c =1/2$
 and the critical exponent of each string order is $\beta=1/8$.
  The phase transitions between Haldane phases
  are a topological phase transition and belong to an Ising universality class.
  Here, the $\theta$ is the interaction parameter from the setting
  $J_y=J \cos\theta$ and $J_z=J\sin\theta$ for the numerical calculation.
  }
 \label{fig1}
\end{figure}
%%%%%%%%%%%%%%%%%%%%%%%%%%%%%%%%%%%%%%%%%%%%%%%%%%%%%%%

 It seems to be believed that the phase transition occurs between two disordered phases
 in the one-dimensional QCM.
 Normally, disordered phases are not characterized by any local order parameter.
 This implies that the phase transition in the one-dimensional QCM
 would not be understood properly within
 the Landau paradigm of spontaneous symmetry breaking~\cite{Sachdev}.
 Consequently,
 such a controversy on the phase transition in the one-dimensional QCM
 would suggest us to consider non-local long-range orders for its proper characterization.
 In order to characterize the phase transition properly, in this paper,
 we investigate non-local string orders in the one-dimensional QCM.
 Actually, a string order as a non-local
 long-range order was introduced by Nijs and Rommelse~\cite{Nijs}
 and Tasaki~\cite{Tasaki},
 and characterizes the Haldane phase in the spin-$1$ Heisenberg chain \cite{Yamamoto}.
 To calculate non-local string orders directly  \cite{Su}, in contrast to an extrapolated value for finite-size lattices, we employ
 the infinite matrix product state (iMPS)~\cite{Vidal1,Vidal2}
 representation with the iTEBD algorithm developed by Vidal~\cite{Vidal2}.
 For a systematic study,
 the second derivative of ground state energy is calculated to reveal
 the phase transitions in the whole interaction parameter range.
 Its singularities indicate that there are
 the four phases separated by the second-order phase transitions.
 We find the four string order parameters that characterize each phases (see in Fig.~\ref{fig1}),
 which means that all the four phases are a topologically ordered phase.
 Furthermore, the critical exponent from the string orders $\beta=1/8$
 and the central charges $c\simeq 1/2$ at the critical points clarify that the topological
 quantum phase transitions (TQPTs) belong to the Ising-type phase transition.
 In addition, the continuous behaviors of the odd- and even-von Neumann entropies
 and the pinch points of the fidelity per lattice site (FLS) verify
 the second-order phase transitions between two topologically ordered phases.

 This paper is organized as follows.
 In Sec.II, we introduce
 the one-dimensional QCM and discuss
 the second derivative of the ground state energy per site.
 In Sec. III, we display string correlations and define properly the four string order
 parameters characterizing the four topologically ordered phases.
 The critical exponents are presented.
 The phase transitions are discussed by employing
 the von Neumann entropy in Sec. IV.
 The TQPTs are classified based on the central charge
 via the finite-entanglement scaling.
 In Sec. V, we discuss the pinch points of the FLS.
 Finally, our conclusion is given in Sec. VI.

\section{Quantum Compass Model and groundstate energy}
 We consider the one-dimensional spin-$1/2$ QCM~\cite{Brzezicki1} written as
\begin{equation}
 H= \sum_{i=-\infty}^\infty \left(
   J_y S^y_{2i-1}\cdot S^y_{2i}
   +
   J_z S^z_{2i}\cdot S^z_{2i+1}
   \right),
 \label{Ham}
\end{equation}
 where $S^y_i$ and $S^z_i$ are the spin-$1/2$ operators on the $i$th site. $J_y$ and
 $J_z$ are nearest-neighbor exchange couplings on the odd and the even bonds, respectively. In
 order to cover the whole range of the parameter $J_y$ and $J_z$, we
 set $J_y=J \cos\theta$ and  $J_z=J \sin\theta$.

%%%%%%%%%%%%%%%%%%%%%%%%%%%%%%%%%%%%%%%%%%%%%%%%%%%%%%%%%
\begin{figure}
\includegraphics [width=0.48\textwidth]{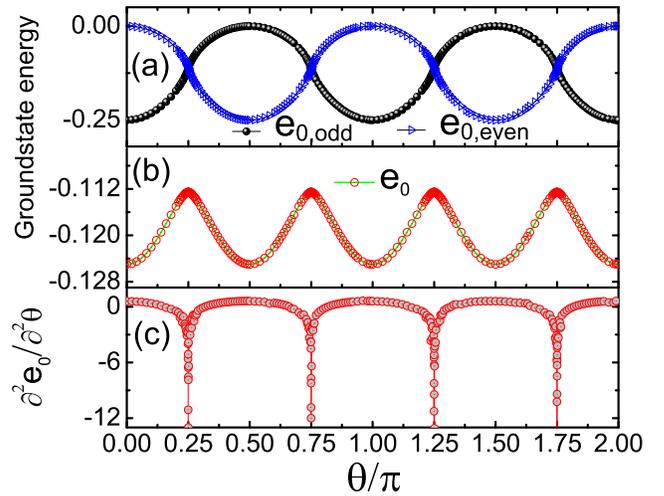}
\caption{ (Color online)
 (a) Groundstate energies per site on odd-/even-bonds $e_{0,odd/even}$,
 (b) average energy $e_0=(e_{0,odd}+e_{0,even})/2$,
 and
 (c) second derivative of the average energy $e_0$
 as a function of the interaction parameter $\theta$.
 Here, the truncation dimension $\chi=40$ is chosen for the iMPS calculation.
 In (c), note that the singular behaviors of the second derivative
 occur at the points $\theta=\pi/4$, $\theta=3\pi/4$,
 $\theta=5\pi/4$, and $\theta=7\pi/4$.
  }
 \label{fig2}
\end{figure}
%%%%%%%%%%%%%%%%%%%%%%%%%%%%%%%%%%%%%%%%%%%%%%%%%%%%%%%%%

 From the iMPS groundstate wavefunction, we obtain the groundstate energy of the QCM.
 In Figs.~\ref{fig2}(a) and \ref{fig2}(b),
 we plot the groundstate energies $e_{0,odd}$ on the odd bond and $e_{0,even}$ on the even bond,
 and the groundstate energy per site $e_0$ as an average value of the energies
 $e_{0,odd}$ and $e_{0,even}$, i.e., $e_0=(e_{0,odd}+e_{0,even})/2$.
 Here, the truncation dimension is chosen as $\chi=40$.
 The energies are shown to be a periodic behavior as a function of the interaction parameter $\theta$.
 One way to know whether there is a phase transition is to check the non-analyticity of the groundstate
 energy on the system parameters.
 Thus,
 in order to see any possible phase transition,
 we calculate the derivative of the energies over
 the interaction parameter $\theta$.
 In the first derivative of the energies over the interaction parameter,
 no singular behavior is noticed in the whole parameter range.
 Then,
 in Fig.~\ref{fig2}(c), we plot the
 second derivative of the energy $e_0$.
 Note that it exhibits the
 singular points at $\theta=\pi/4$, $\theta=3\pi/4$,
 $\theta=5\pi/4$, and $\theta=7\pi/4$.
 This result means that, at the singular points,
 the quantum phase transitions occur and
 they are of the second-order.
 As we introduced the controversy of the phase transition in the QCM,
 the critical point $\theta=\pi/4$ in our calculation corresponds to the
 the critical point $J_y=J_z$ investigated in previous studies.
 Consequently, our second derivative of the groundstate energy
 shows that the phase transition in the QCM should be of the second-order.
 Moreover, the critical lines separate the parameter space into the four regions
 [denoted by I, II, III, and IV in Fig. \ref{fig1}],
 which may indicate four possible phases.
 Then, in order to characterize the four possible phases,
 we discuss string correlations in the next section.

\section{string order parameters and topological quantum phase transitions}
 The QCM has the different strengthes of the spin exchange interaction
 depending on the odd and the even bonds.
 One can then define string correlations based on the bond alternation \cite{Hida,Cho}.
 Let us first consider the string correlations defined as
\begin{subequations}
 \begin{eqnarray}
 %\nonumber
 O^\alpha_{s,odd} \left( 2i-1, 2j \right)
 \!\!\! &=&  \!\!\!
  \left\langle
     S^\alpha_{2i-1} \exp \left[i\pi\sum_{k=2i}^{2j-1}S^\alpha_k\right] S^\alpha_{2j}
  \right\rangle
 \label{Ostring}
   \\
 O^\alpha_{s,even} \left( 2i, 2j+1 \right)
 \!\!\! &=&  \!\!\!
  \left\langle
  S^\alpha_{2i} \exp\left[i\pi\sum_{k=2i+1}^{2j}S^\alpha_k \right] S^\alpha_{2j+1}
  \right\rangle ,
 \label{Estring}
\end{eqnarray}
\end{subequations}
where $\alpha=x$, $y$, and $z$.
 We observe numerically that the $x$ components of
 the string correlations $O^{x}_{str,odd/even}$
 decrease to zero within the lattice distance $|i-j|=6$
 in the whole parameter range.
%

%%%%%%%%%%%%%%%%%%%%%%%%%%%fig 3%%%%%%%%%%%%%%%%%%%%%%%%%%%%%%%%
\begin{figure}
 \includegraphics[width=0.3\textwidth] {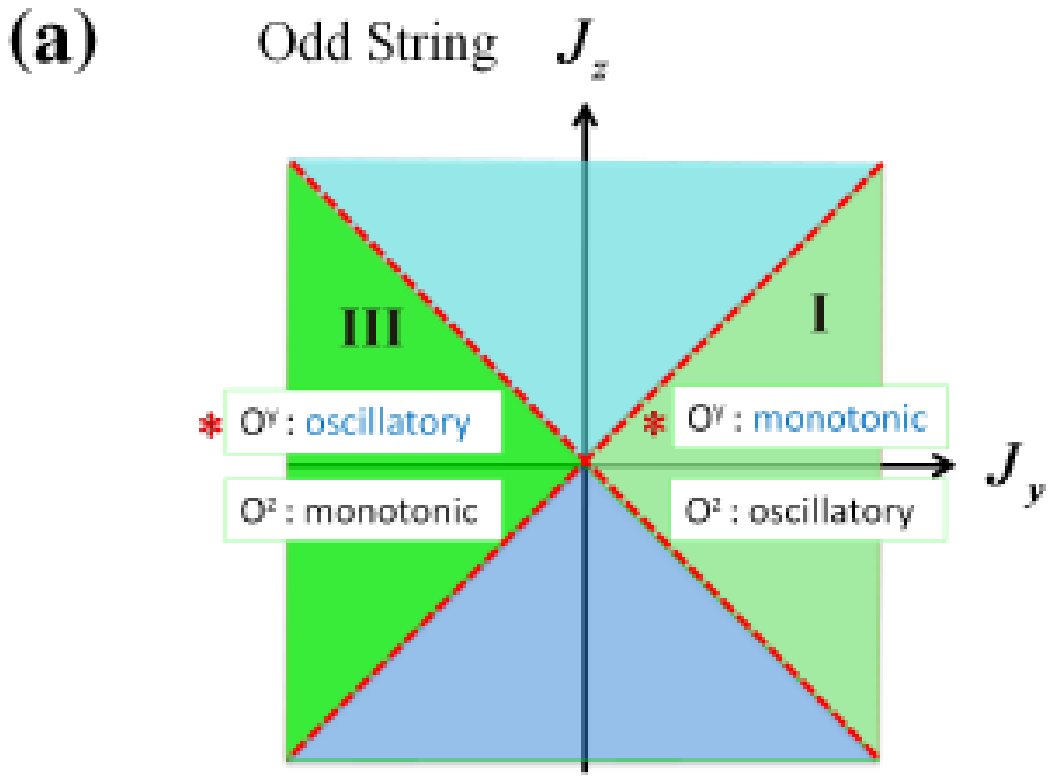}
 \includegraphics[width=0.4\textwidth] {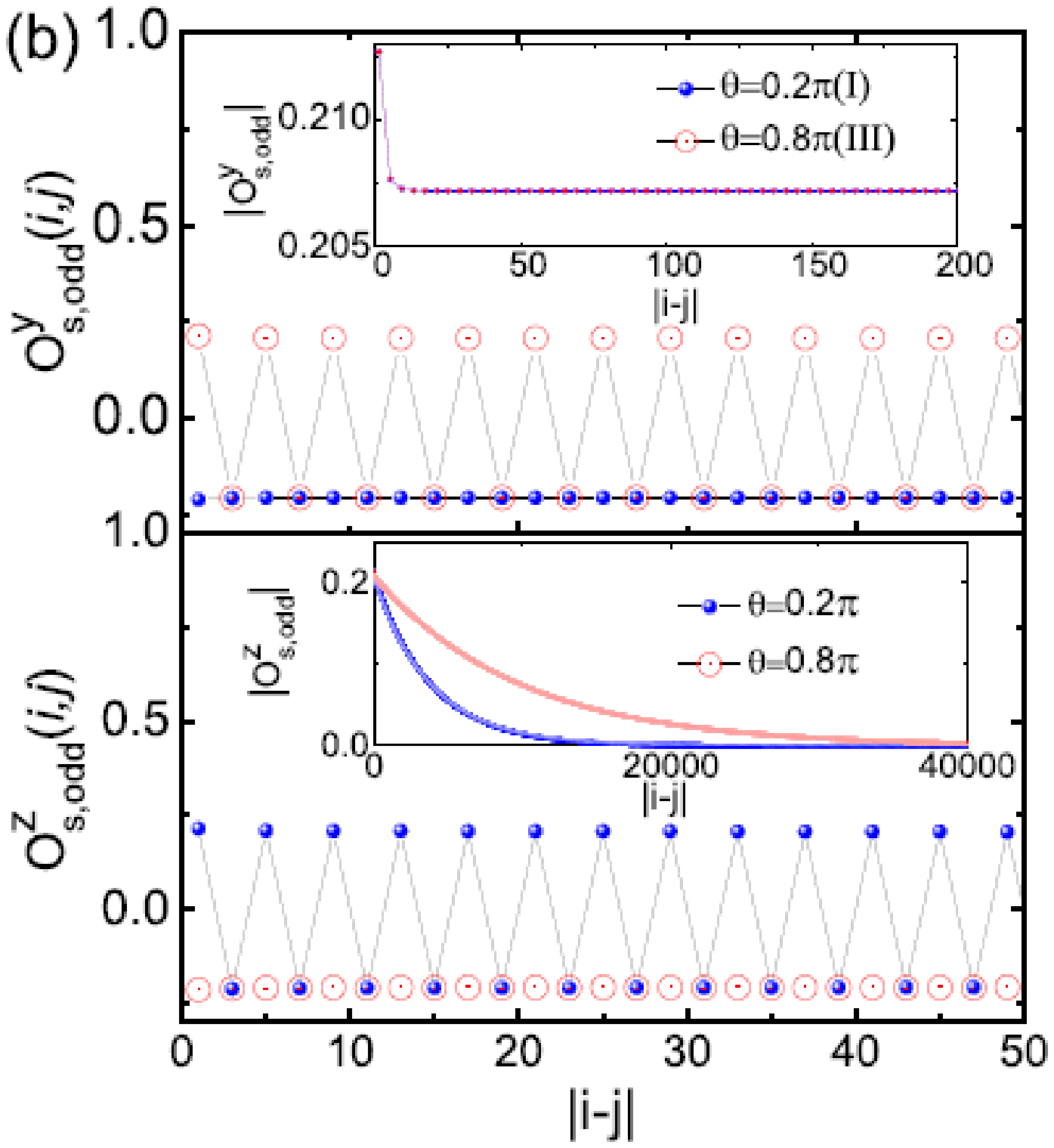}
 \caption{ (Color online)
 (a) Behaviors of the odd string order parameters (indicated by
   the asterisk)  in $J_y$-$J_z$ plane.
 (b) String correlations $O^{y/z}_{str,odd}$ for $\theta=0.2\pi$ and $\theta=0.8\pi$.
 In the insets,
 note that $O^{y}_{str,odd}$'s are saturated to a finite value,
 while $O^{z}_{str,odd}$'s decay to zero for very large distance.
 }
 \label{fig3}
\end{figure}
%%%%%%%%%%%%%%%%%%%%%%%%%%%%fig 3%%%%%%%%%%%%%%%%%%%%%%%%%%%%%%%%

 {\it Behaviors of odd string correlations.}$-$
 In Fig. \ref{fig3}(a), we summarize the short- and long-distance behaviors of
 the odd string correlations $O^{y}_{str,odd}$ in $J_y$-$J_z$ plane.
 (i) For $|J_y| < |J_z|$ (the regions II and IV in Fig. \ref{fig1}),
 the odd string correlations $O^{y/z}_{str,odd}$
 decrease to zero within the lattice distance $|i-j|=80$.
 (ii) For $|J_y| > |J_z|$ (the regions I and III in Fig. \ref{fig1}),
 the absolute value of $O^{y}_{str,odd}$ are saturated to a finite value
 while $O^{z}_{str,odd}$ decays to zero very slowly, which means
 $O^{y}_{str,odd}$ as a non-local long-range order parameter
 [indicated by the asterisk  in Fig. \ref{fig3}(a)]
 can characterize a topologically ordered phase.
 Further, if $J_y > 0$[region I] ($J_y < 0$ [region III]),
 $O^{y}_{str,odd}$ shows a monotonic (oscillatory) saturation
 and $O^{z}_{str,odd}$ displays  an oscillatory (monotonic) decaying to zero.

 As an example, in Fig. \ref{fig3}(b), we plot
 the odd string correlations $O^{y/z}_{str,odd}$
 as a function of the lattice distance $|i-j|$
 for $\theta=0.2\pi$ (the range I)
 and $\theta=0.8\pi$  (the region III).
 The string correlations are shown a very distinct behavior.
 For $\theta=0.2\pi$, the $O^{y}_{str,odd}$ has a minus sign, while the $O^{z}_{str,odd}$
 has an alternating sign depending on the lattice distance.
 In contrast to the case of $\theta=0.2\pi$, for $\theta=0.8\pi$,
 the $O^{z}_{str,odd}$ has a minus sign, while the $O^{y}_{str,odd}$
 has an alternating sign depending on the lattice distance.
 From the short-distance behaviors, as shown in Fig. \ref{fig3}(b),
 it is hard to see whether the string correlations decay to survive
 in the long distance limit (i.e., $|i-j| \rightarrow \infty$).
 In order to study the correlations in the limit of
 the infinite distance,
 one can then set a truncation error $\varepsilon$ rather
 than the lattice distance, i.e., $O(|i-j|)-O(|i-j-1|) < \varepsilon$.
 In this study, for instance, $\varepsilon=10^{-8}$ is set.
 The insets of Fig. \ref{fig3}(b) show
 the string correlations for relatively very large lattice distance.
 We see clearly that the $O^{y}_{str,odd}$'s have a finite value
 while the the $O^{z}_{str,odd}$'s decay to zero
 (around the lattice distance $|i-j|\sim 5\times 10^{4}$).
 As a consequence, the parameter regions I and III can be characterized by
 the odd string long-range order parameters.
 As discussed above, the odd string correlations have the two characteristic behaviors,
 i.e., one is a monotonic saturation for $\theta=0.2\pi$,
 the other is an oscillatory saturation for $\theta=0.8\pi$.
 Such a distinguishable behavior of the string correlations
 allows us to say that the region I ($J_y > 0$) and the region III ($J_y < 0$)
 are a different phase each other and we call
 the \textit{monotonic} odd string order and the \textit{oscillatory} odd string order, respectively.

%%%%%%%%%%%%%%%%%%%%%%%%%%%fig 4%%%%%%%%%%%%%%%%%%%%%%%%%%%%%%%%
\begin{figure}
 \includegraphics[width=0.3\textwidth] {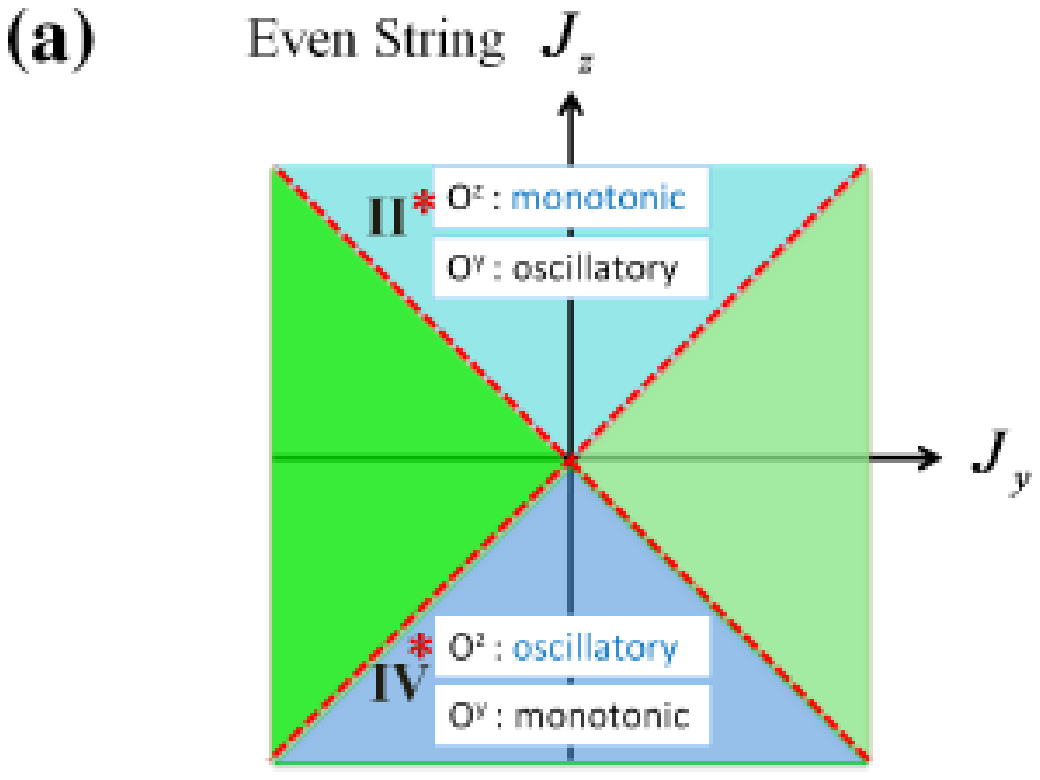}
 \includegraphics[width=0.4\textwidth] {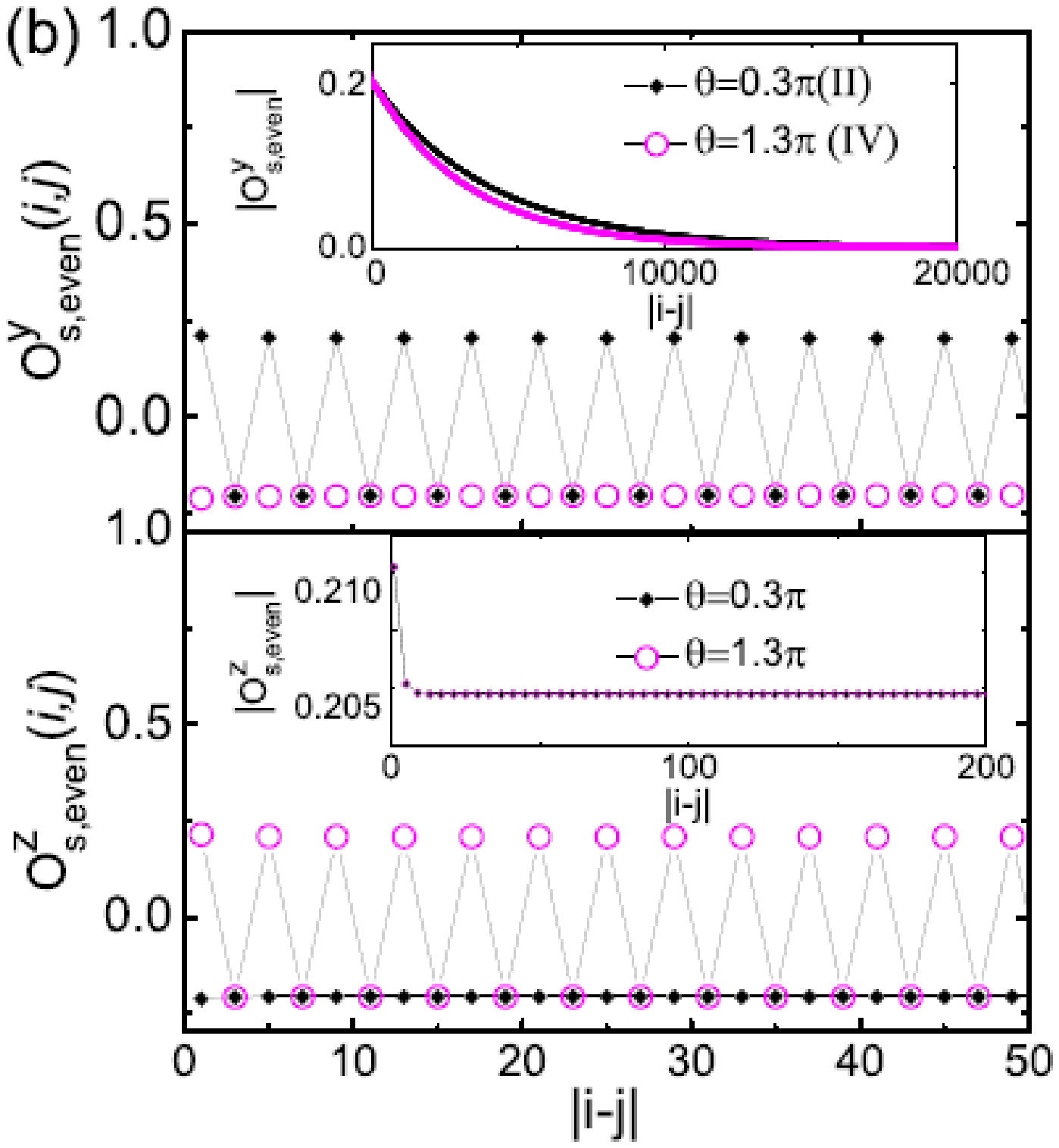}
 \caption{ (Color online)
 (a) Behaviors of the even string order parameters (indicated by
   the asterisk) in $J_y$-$J_z$ plane.
 (b) String correlations $O^{y/z}_{str,even}$ for $\theta=0.3\pi$ and $\theta=1.3\pi$.
 In the insets,
 note that $O^{z}_{str,even}$'s are saturated to a finite value,
 while $O^{y}_{str,even}$'s decay to zero for very large distance.
 }
 \label{fig4}
\end{figure}
%%%%%%%%%%%%%%%%%%%%%%%%%%%%fig 4%%%%%%%%%%%%%%%%%%%%%%%%%%%%%%%%

 {\it Behaviors of even string correlations.}$-$
 Similarly to the odd string correlations,
 the even string correlations show the two characteristic behaviors.
 In Fig. \ref{fig4}(a), we summarize the short- and long-distance behaviors of
 the even string correlations $O^{y}_{str,even}$ in $J_y$-$J_z$ plane.
 (i) For $|J_z| < |J_y|$ (the regions I and III in Fig. \ref{fig1}),
 the even string correlations $O^{y/z}_{str,even}$
 decrease to zero within the lattice distance $|i-j|=80$.
 (ii) For $|J_z| > |J_y|$ (the regions II and IV in Fig. \ref{fig1}),
 the absolute value of $O^{z}_{str,even}$ are saturated to a finite value
 while $O^{y}_{str,even}$ decays to zero very slowly, which means
 $O^{z}_{str,even}$ as a non-local long-range order parameter
 [indicated by an asterisk  in Fig. \ref{fig4}(a)]
 characterizes a topologically ordered phase.
 Further, if $J_z > 0$[region II] ($J_z < 0$ [region IV]),
 $O^{z}_{str,even}$ shows a monotonic (oscillatory) saturation
 and $O^{y}_{str,even}$ displays an oscillatory (monotonic) decaying to zero.

 As an example, in Fig. \ref{fig4}(b), we plot
 the even string correlations $O^{y/z}_{str,even}$
 as a function of the lattice distance $|i-j|$
 for $\theta=0.3\pi$ (the range II)
 and $\theta=1.3\pi$  (the region IV).
 For $\theta=0.3\pi$, the
 $O^{y}_{str,even}$ has an alternating sign depending on the lattice distance,
 while the $O^{z}_{str,even}$ has a minus sign.
 In contrast to the case of $\theta=0.3\pi$, for $\theta=1.3\pi$,
 the $O^{z}_{str,even}$ has an alternating sign depending on the lattice distance,
 while the $O^{y}_{str,even}$ has a minus sign.
 Similarly to the odd string correlations,
 the short distance behaviors of the even string correlations
 show to the difficulty to see which the string correlations survive in the long distance limit (i.e., $|i-j| \rightarrow \infty$).
 By using the truncation error  $\varepsilon=10^{-8}$,
 we plot the string correlations for relatively very large lattice distance
 in the insets of Fig. \ref{fig4}(b).
 We see clearly that the $O^{z}_{str,even}$'s have a finite value
 while the the $O^{y}_{str,odd}$'s decay to zero
 (around the lattice distance $|i-j|\sim 2\times 10^{4}$).
 As a result, the parameter regions II and IV can be characterized by
 the even string long-range order parameters.
 The even string correlations also have the two characteristic behaviors,
 i.e., one is a monotonic saturation for $\theta=0.3\pi$,
 the other is an oscillatory saturation for $\theta=1.3\pi$.
 The region II ($J_z > 0$) and the region IV ($J_z < 0$)
 are a different phase each other and we call
 the \textit{monotonic} even string order and the \textit{oscillatory} even string order, respectively.

%
%%%%%%%%%%%%%%%%%%%%%%%%%%%fig 5%%%%%%%%%%%%%%%%%%%%%%%%%%%%%%%%
\begin{figure}
 \includegraphics[width=0.45\textwidth] {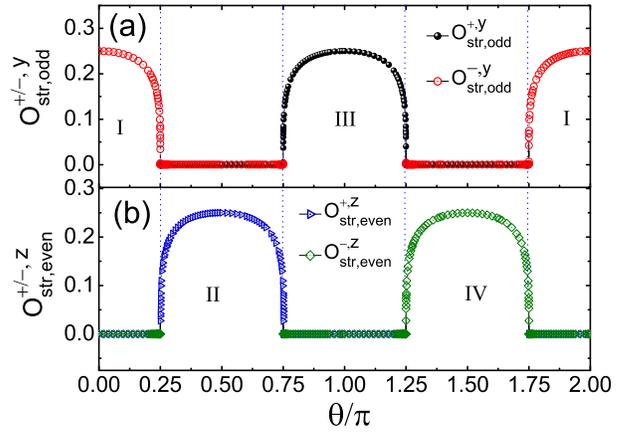}
 \caption{ (Color online)
 String order parameters (a) $O^{+/-, y}_{str,odd}$ and (b) $O^{+/-, z}_{str,even}$  as a function of $\theta$. The order parameters are defined in the text.}
 \label{fig5}
\end{figure}
%%%%%%%%%%%%%%%%%%%%%%%%%%%%fig 5%%%%%%%%%%%%%%%%%%%%%%%%%%%%%%%%
%
%
 {\it Phase diagram from string order parameters.}$-$
 As we discussed, the even and odd string correlations have shown two characteristic behaviors, i.e.,
 one is monotonic, the other is oscillatory.
 Then, one may define a proper long-range order
 based on the behaviors of the odd and the even string correlations.
 We define the long-range string order parameters as follows:
\begin{subequations}
 \begin{eqnarray}
 %\nonumber
 O^{+, y}_{str,odd}  \!\!\! &=&  \!\!\! - \lim_{|i-j|\rightarrow\infty}
  O^y_{s,odd} \left( 2i-1, 2j \right), \\
 O^{-, y}_{str,odd}  \!\!\! &=&  \!\!\! - \lim_{|i-j|\rightarrow\infty}
 (-1)^{(j-i+1)}  O^y_{s,odd} \left( 2i-1, 2j \right), \\
 O^{+, z}_{str,even}  \!\!\!  &=& \!\!\! - \lim_{|i-j|\rightarrow\infty}
 O^z_{s,even} \left( 2i, 2j+1 \right),   \\
 O^{-,  z}_{str,even}  \!\!\!  &=&  \!\!\! -  \lim_{|i-j|\rightarrow\infty}
 (-1)^{(j-i+1)} O^z_{s,even} \left( 2i, 2j+1 \right),
 \label{equ:strorder}
\end{eqnarray}
\end{subequations}
 where, actually, the superscript $+$ ($-$)
 of the string order parameters denotes the monotonic behavior (the oscillatory behavior).

 The defined string orders are calculated from the iMPS groundstate wave function.
 In Fig.~\ref{fig5}, we display
 the string order parameters as a function of the interaction parameter $\theta$.
 In Fig.~\ref{fig5}(a), it is clearly shown that
 the odd string order parameters are finite for the region I ($ -\pi/4< \theta < \pi/4$)
 and the region III ($ 3\pi/4< \theta < 5\pi/4$).
 Further, the monotonic odd string order parameter $O^{+, y}_{str,odd}$
 characterizes the region I and the oscillatory odd string order parameter $O^{-, y}_{str,odd}$
 does the region III.
 Similarly to the odd string order parameters,
 the $z$ components of the even string order parameters $O^{+, z}_{str,even}$ and $O^{-, z}_{str,even}$  are finite for the region II ($ \pi/4< \theta < 3\pi/4$)
 and the region IV ($ 5\pi/4< \theta < 7\pi/4$).
 The monotonic even string order parameter $O^{+, z}_{str,even}$
 characterizes the region II and the oscillatory even string order parameter $O^{-, y}_{str,even}$
 does the region IV.
 Consequently,
 the four regions in $J_y$-$J_z$ plane [Fig. \ref{fig1}] are characterized by
 the four string order parameters $O^{+/-, y}_{str,odd}$ and
 $O^{+/-, z}_{str,even}$, respectively,
 which implies that a different hidden $Z_2\times Z_2$ breaking symmetry occurs in each phase.
 Therefore, the one-dimensional QCM has the four distinct topologically ordered phases rather than disordered phases suggested in previous studies.
 The system undergoes a topological quantum phase transition between two
 topological ordered phases as the interaction parameter crosses the critical lines
 $|J_y|=|J_z|$.
 In addition, the continuous behaviors of the string order parameters across the critical lines
 show that the
 topological quantum phase transitions are of the continuous (second-order) phase transition rather than
 the discontinues (first-order) phase transition.

 In a previous study \cite{Liu1} on an EQCM,
 the existence of a string order has been noticed numerically for a relevant interaction parameter range.
 However, any characterization of phase has not been made in association with the one-dimensional
 QCM.
 However, the one-dimensional spin-$1/2$ Kitaev model \cite{Kitaev}, which is equivalent to
 the one-dimensional QCM, has shown to have
 two string order parameters \cite{Feng}
 based on the dual spin correlation function \cite{Pfeuty}
 by using a dual transformation
 \cite{Fradkin,Kohmoto} mapping the model into a one-dimensional Ising model with a transverse field.
 The actual parameter range in the one-dimensional Kitaev model studied in Ref. \onlinecite{Feng}
 corresponds to $J_y > 0$ and $J_z > 0$ in our one-dimensional QCM.
 The system was discussed to undergo
 a topological quantum phase transition at the critical point $J_y=J_z > 0$ ($\theta=\pi/4$).
 In this sense, in the case of $J_y, J_z > 0$ in the one-dimensional QCM,
 we have numerically demonstrated and verified the existence of the string order parameters
 and the topological quantum phase transition as discussed in Ref. \onlinecite{Feng}.
%

%
%%%%%%%%%%%%%%%%%%%%%%%%%%%fig 6%%%%%%%%%%%%%%%%%%%%%%%%%%%%%%%%
\begin{figure}
 \includegraphics[width=0.48\textwidth] {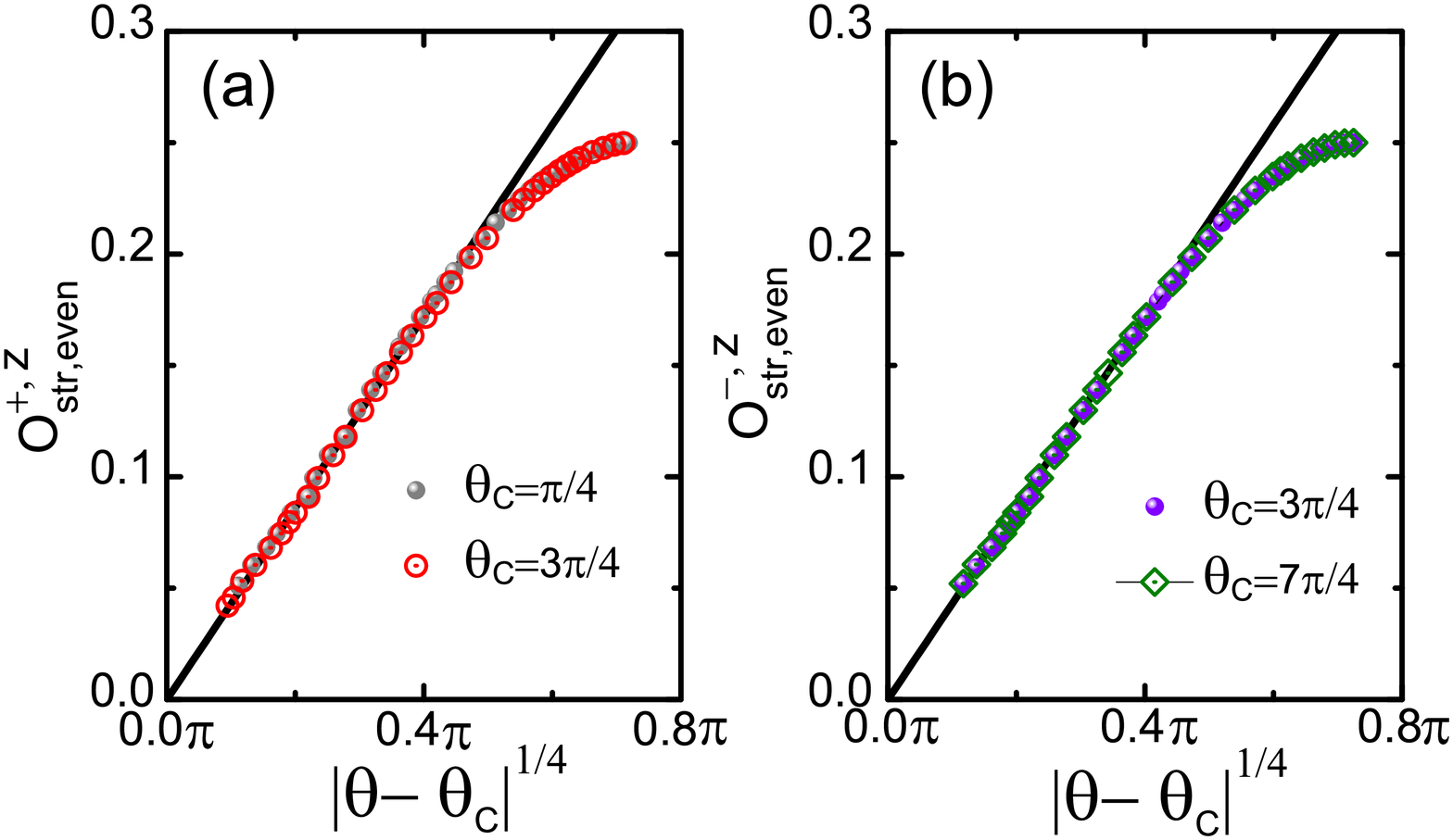}
 \caption{ (Color online)
 String order parameters (a) $O^{+, z}_{str,even}$ and (b)
 $O^{-, z}_{str,even}$  as a function of $|\theta-\theta_c|^{1/4}$ for
 $\theta_c=\pi/4$, $3\pi/4$, $5\pi/4$, and $7\pi/4$.
 }
 \label{fig6}
\end{figure}
%%%%%%%%%%%%%%%%%%%%%%%%%%%%fig 6%%%%%%%%%%%%%%%%%%%%%%%%%%%%%%%%

 {\it Critical exponents.}$-$
 In the critical regimes, as the order parameters, the string orders should show a scaling behavior to
 characterize the phase transitions.
 We plot the string order parameters $O^{+, z}_{str,even}$ [Fig. \ref{fig6}(a)]
 and $O^{-, z}_{str,even}$ [Fig. \ref{fig6}(b)]
 as a function of $|\theta-\theta_c|^{1/4}$ with the critical points $\theta_c=\pi/4$, $3\pi/4$,
 $5\pi/4$, and $7\pi/4$.
 It is shown that all the string order parameters
 nearly collapse onto one scaling fitting function in the critical regimes, i.e.,
 they scales as $O^{\pm, z/y}_{str,even/odd} \propto |\theta-\theta_c|^{1/4}$.
 As a result, the same critical exponents are given
 as $\beta=1/8$ via $O^{+/-, z}_{str,even} \propto |\theta-\theta_c|^{2\beta}$ \cite{Shelton},
 which reveals that the TQPTs belong the Ising-type phase transition.

\section{Entanglement entropy and central charge}
 Quantum entanglement in many-body systems
 can be quantified by the von Neumann entropy that
 is a good measure of bipartite entanglement between two subsystems
 of a pure state\cite{Osterloh,Amico}.
 Generally, for one-dimensional quantum spin lattices,
 at critical points,the von Neumann entropy exhibits its logarithmic
 scaling conforming conformal invariance.
 Its scaling is governed by a universal factor, i.e., a central charge $c$ of the associated
 conformal field theory. The central charge allows us to classify
 a universality class~\cite{Cardy} of quantum phase transition.
 In our iMPS representation,
 a diverging entanglement at quantum critical points gives simple
 scaling relations for (i) the von Neumann entropy $S$ and
(ii) a correlation length $\xi$ with respect to the truncation
 dimension $\chi$ ~\cite{Tagliacozzo} as follows:
 \begin{subequations}
 \begin{eqnarray}
 %\nonumber
 \xi (\chi) & \propto& \xi_0 \chi^{\kappa}
 \label{equ:SS}  \\
  S (\chi) & \propto & \frac{c \kappa}{6} \log_2{\chi},
 \label{equ:CC}
\end{eqnarray}
\end{subequations}
 where $\kappa$ is a so-called finite-entanglement scaling exponent and $\xi_0$ is a constant.
 Thus, one can calculate a central charge by using Eqs. (\ref{equ:SS}) and (\ref{equ:CC}).

 In order to obtain the von Neumann entropy,
 we partition the spin chain into the two parts denoted by the
 left semi-infinite chain $L$ and the right semi-infinite chain $R$.
 In terms of the reduced density matrix $\varrho_L$ or $\varrho_R$ of
 the subsystems $L$ and $R$,
 the von Neumann entropy can be defined as
 $S=-\mathrm{Tr}\varrho_L\log_2\varrho_L=-\mathrm{Tr}\varrho_R\log_2\varrho_R$.

 In the iMPS representation,
 the iMPS groundstate wavefunction can be written by
 the Schmidt decomposition
 $|\Psi\rangle=\sum_{\alpha=1}^{\chi}\lambda_{\alpha} |\phi^L_\alpha\rangle|\phi^R_\alpha\rangle$,
 where $|\phi^L_\alpha\rangle$ and $|\phi^R_\alpha\rangle$
 are the Schmidt bases for the semi-infinite chains
 $L(-\infty,\cdots,i)$ and $R(i+1, \cdots,\infty)$, respectively.
 $\lambda_{\alpha}^2$ are actually
 eigenvalues of the reduced density matrices for the two
 semi-infinite chains $L$ and $R$.
 In our four-site translational
 invariant iMPS representation, we have the four Schmidt coefficient
 matrices $\lambda_A$, $\lambda_B$, $\lambda_C$ and $\lambda_D$,
 which means that there are the four possible ways
 for the partitions.
 Due to the two-site translational invariance of the QCM,
 in fact, we have $\lambda_A=\lambda_C$ and $\lambda_B=\lambda_D$, i.e.,
 one partition is on the odd sites, the other is on
 the even sites.
 From the $\lambda_{even}$ and $\lambda_{odd}$,
 one can obtain the two von Neumann entropies depending on the
 odd- or even-site partitions as
\begin{equation}
 S_{even/odd}=-\sum_{\alpha=1}^\chi \lambda_{even/odd,\alpha}^2 \log_2 \lambda_{even/odd,\alpha}^2,
 \label{entropy}
\end{equation}
 where $\lambda_{even/odd,\alpha}$'s are diagonal elements of the matrix
 $\lambda_{even/odd}$.

  %%%%%%%%%%%%%%%%%%%%%%%%%%%%%%%%%%%%%%%%%%%%%%%%%%%%%%%%%%%%%%%
\begin{figure}
  \includegraphics[width=0.46\textwidth]{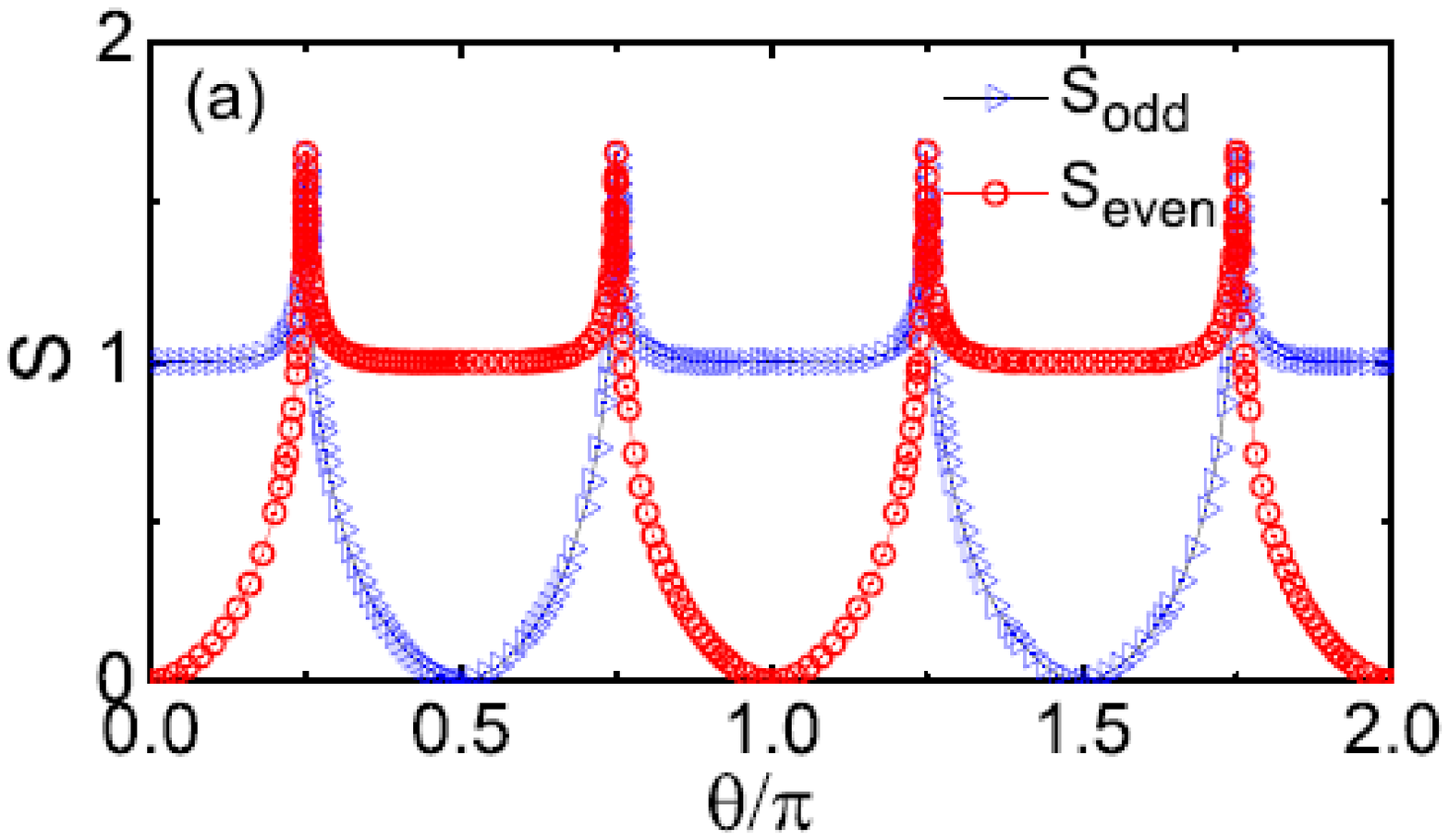}
  \includegraphics[width=0.46\textwidth]{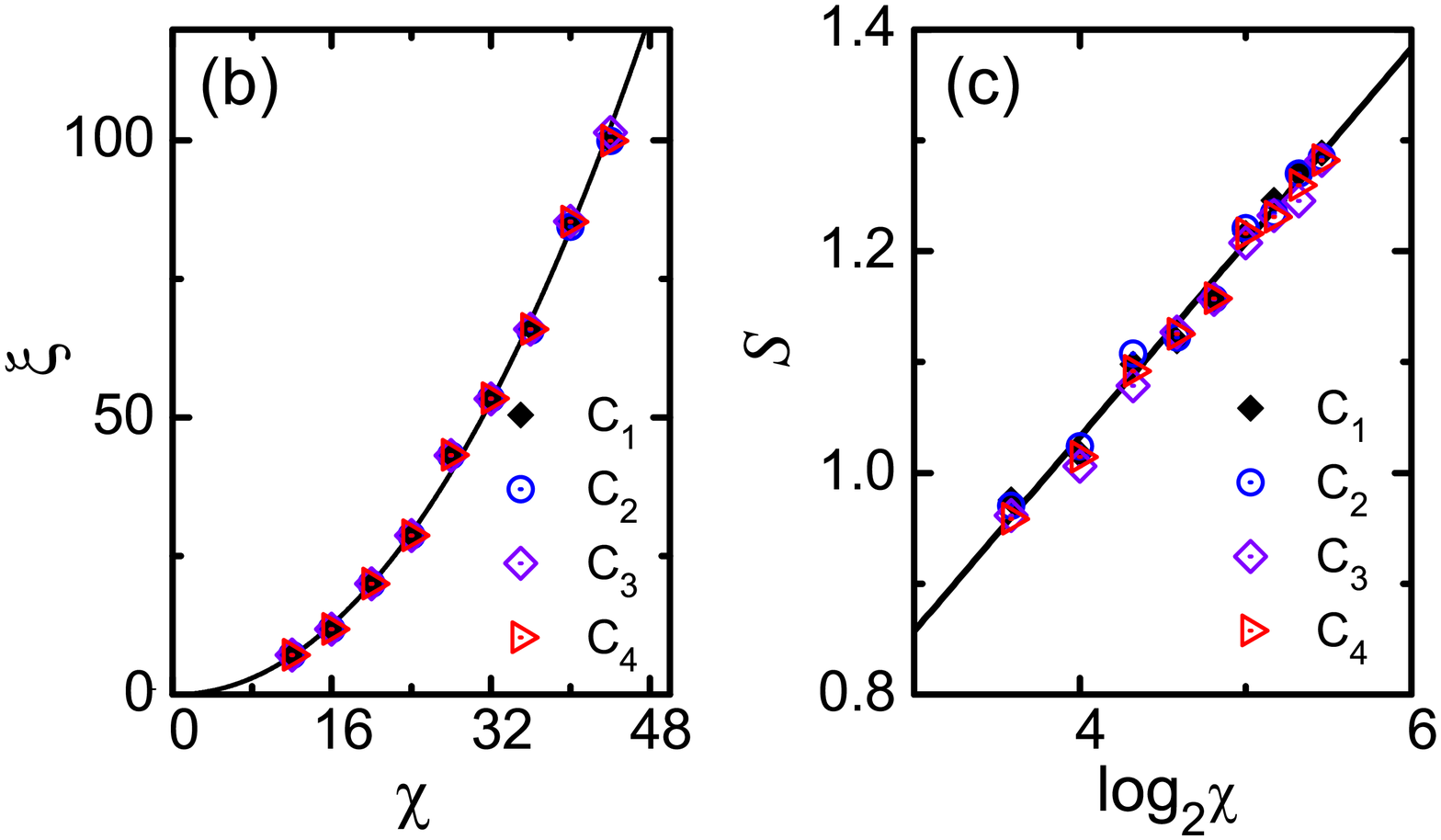}
 \caption{(Color online) Von Neumann entropies $S_{odd}$ and
 $S_{even}$ as a function of the interaction parameter $\theta$.
 Note that the entropy singular points
  at $\theta=\pi/4$, $3\pi/4$,
 $5\pi/4$, and $7\pi/4$ correspond to the critical points from
 the string order parameters.
 (b) Correlation length $\xi(\chi)$ as a function of the
  truncation dimension $\chi$ at the critical points $C_1(J_x, J_y)=(1, 1)$, $C_2=(-1, 1)$,
   $C_3=(-1, -1)$, and $C_4=(1, -1)$.
 (c) Von Neumann entropy $S(\chi)$ as a function of $\chi$ at the critical points.
} \label{fig7}
\end{figure}
%%%%%%%%%%%%%%%%%%%%%%%%%%%%%%%%%%%%%%%%%%%%%%%%%%%

  In Fig.~\ref{fig7}(a),
  we plot
  the von Neumann entropies $S_{odd}(\theta)$ and $S_{even}(\theta)$
  as a function of the control parameter $\theta$.
  One can easily notice that
  there are the four singular points
  $\theta=\pi/4$, $3\pi/4$,
  $5\pi/4$, and $7\pi/4$ in both the odd-bond and the even-bond entropies.
 The four singular points of the von Neumann entropies
 indicate a quantum phase transition at those points.
 It should be noted that the detected transition points from the von Neumann entropies
 correspond to the critical points
 from the second derivative of the groundstate energy and the string order parameters.
 The continuous behaviors of von Neumann entropies around critical points
 also indicate the occurrence of the continuous (second-order) quantum phase transition
 as the system crosses the transition points.
 Hence, it is shown that the von Neumann entropy can detect the topological quantum phase
 transitions.

  In Figs.~\ref{fig7}(b) and \ref{fig7}(c),
  we plot the correlation length $\xi(\chi)$ as a function of the truncation dimension $\chi$
  and the von Neumann entropy $S(\chi)$ as a function of  $\chi$ at the critical points
  $C_1(J_1,J_2)=(1, 1)$, $C_2=(-1, 1)$, $C_3=(-1, -1)$, and $C_4=(1, -1)$, respectively.
 The truncation dimensions are taken as $\chi=12, 16, 20, 24, 28, 32, 40$, and $44$.
 The correlation length $\xi(\chi)$ and the von Neumann entropy $S(\chi)$
 diverge as the truncation dimension $\chi$ increases.
 Using the numerical fitting function $\xi(\chi)=\xi_0 \chi^{\kappa}$ in Eq. (\ref{equ:SS}),
 the fitting constants are obtained as
 (i) $\xi_0=0.04$ and $\kappa=2.071$ at  $C_1$,
 (ii) $\xi_0=0.041$ and $\kappa=2.068$ at $C_2$,
 (iii) $\xi_0=0.039$ and $\kappa=2.087$ at $C_3$, and
 (iv) $\xi_0=0.041$ and $\kappa=2.065$ at $C_4$.
  In order to obtain the central charge, we use the numerical fitting function of
  the von Neumann entropy $S(\chi)=(c\kappa/6)\log_2 \chi + S_0$.
  As shown in Figs.~\ref{fig7}(c),
  the linear scaling behaviors of the entropies
  give
 (i) $c=0.5079$ with $S_0=0.331$ at $C_1$,
 (ii) $c=0.4992$ with $S_0=0.3464$ at $C_2$,
 (iii) $c=0.5048$ with $S_0=0.314$ at $C_3$, and
  (iv) $c=0.5983$ with $b=0.34$ at $C_4$.
  Our central charges are very close to the value $c=0.5$, respectively.
 Consequently,
 the topological quantum phase transitions at all the critical points
 belong to the same universality class, i.e., the Ising universality class.
 This result is consistent with the universality class
 from the critical exponent $\beta=1/8$ of the string order parameters.

 \section{Fidelity per lattice site}
%%%%%%%%%%%%%%%%%%%%%%%%%%%%fig 3%%%%%%%%%%%%%%%%%%%%%%%%%%%%%%%%
\begin{figure}
 \includegraphics[width=0.45\textwidth] {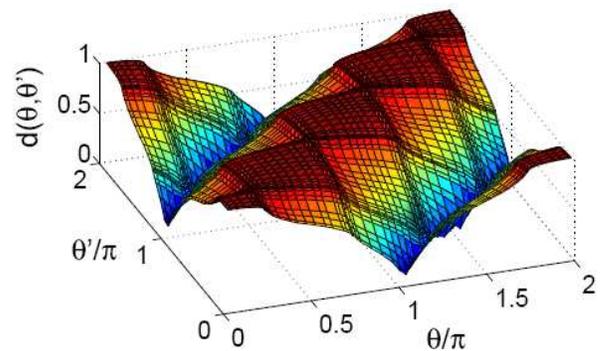}
 \caption{ (Color online)
 Fidelity per site $d(\theta,\theta')$ surface as a
 function of the two parameters $\theta$ and $\theta'$.
 The pinch points $\theta=\pi/4$, $3\pi/4$,
 $5\pi/4$, and $7\pi/4$ on the FLS surface
 indicate the occurrence of the continuous phase transitions.
 }
 \label{fig8}
\end{figure}
%%%%%%%%%%%%%%%%%%%%%%%%%%%%%fig 3%%%%%%%%%%%%%%%%%%%%%%%%%%%%%%%%

 Similarly to the von Neumann entropy,
 the fidelity per lattice site (FLS) \cite{Zhou}
 is known to enable us to detect a phase transition point
 as an universal indicator without knowing
 any order parameters.
 From our iMPS groundstate wave function $|\Psi(\theta)\rangle$ with the interaction parameter $\theta$,
 we define the fidelity as $F(\theta,\theta')= |\langle\Psi(\theta)| \Psi(\theta')\rangle|$.
 Following Ref. \onlinecite{Zhou},
 the ground-state FLS $d(\theta,\theta')$ can then be defined as
\begin{equation}
 \ln d(\theta,\theta') = \lim_{L \rightarrow \infty}
                         \frac {\ln F(\theta,\theta')}{L},
 \label{dinf}
\end{equation}
 where $L$ is the system size.

  In Fig.~\ref{fig8},
  the groundstate FLS $d(\theta,\theta')$ is plotted in $\theta$-$\theta'$ parameter space.
  The FLS surface reveals that there are the four
  pinch points $\theta=\pi/4$, $3\pi/4$, $5\pi/4$, and $7\pi/4$.
  Each pinch point corresponds to each phase transition point
  from the second-order derivative of the ground-state energy, the string order parameters,
  and the von Neumann entropy.
  In addition,
  the continuous behavior of the groundstate FLS
  verifies that the second-order quantum phase transitions occur at the pinch points.

\section{Conclusion}
 We have investigated the quantum phase transition in the one-dimensional QCM
 by using the iMPS representation with the iTEBD algorithm.
 To characterize quantum phases in the one-dimensional QCM,
 we introduced the odd and the even string correlations
 based on the alternating strength of the exchange interaction.
 We have observed that there are the two distinct behaviors of the odd and the string correlations,
  i.e., one is of the monotonic, (ii) the other is of the oscillatory.
 Based on the topological characterization, we find that there are the four
 topologically ordered phases in the whole interaction parameter range [Fig. \ref{fig1}].
 In the critical regimes, the critical exponents of the string order parameters
 are obtained as $\beta =1/8$, which implies that the topological quantum phase transitions
 belong to the Ising type of universality class.
 Consistently, we obtain the central charges $c=1/2$ from the entanglement entropy.
 In addition,
 the singular behaviors
 of the second-order derivatives of ground state energy,
 the string order parameters characterizing the four Haldane phases,
 the continues behaviors of the von Neumann entropy and the FLS
 allow us to conclude that the phase transitions in the one-dimensional QCM
 are of the second-order, in contrast to previous studies.

\begin{acknowledgments}
 We thank Huan-Qiang Zhou for useful comments.
 HTW acknowledges a support by the National Natural Science Foundation of
 China under the Grant No. 11104362.
 The work was supported by the National Natural Science Foundation of
 China under the Grants No. 11374379.
\end{acknowledgments}

\end{document}